\documentclass{aa}

\usepackage[utf8]{inputenc}
\usepackage{graphicx}
\usepackage{natbib}
\usepackage{comment}
\usepackage{hyperref}
\hypersetup{
  colorlinks   = true, 
  urlcolor     = red, 
  linkcolor    = blue, 
  citecolor   = blue 
}
\usepackage{array}
\usepackage{makecell}
\usepackage[table,xcdraw]{xcolor}
\usepackage{multirow}
\usepackage{amsmath}

\begin{document}

\title{The influence of different coronal hole geometries on simulations of coronal wave -- coronal hole interaction}

\titlerunning{Geometrical properties of CW-CH interaction}
\authorrunning{Piantschitsch, I., Terradas, J. et al.}

\author{Piantschitsch, I.$^{1,2,3}$, Terradas, J.$^{1,2}$, Soubrie, E.$^{2,4}$, Heinemann, S.G.$^5$, Hofmeister S.J.$^6$, Soler, R.$^{1,2}$, and Temmer, M.$^3$}

\institute{$^1$Departament de F\'\i sica, Universitat de les Illes Balears (UIB),
E-07122, Spain \\   
$^2$Institute of Applied Computing \& Community Code (IAC$^3$),
UIB, Spain\\
$^3$Institute of Physics, University of Graz, Universit\"atsplatz 5, A-8010 Graz, Austria\\
$^4$Institut d'Astrophysique Spatiale, CNRS, Univ. Paris-Sud, Universit\'e Paris-Saclay, B\^at. 121, 91405 Orsay, France
\\
$^5$Department of Physics, University of Helsinki, P.O. Box 64, 00014, Helsinki, Finland
\\
$^6$Leibniz Institute for Astrophysics Potsdam, An der Sternwarte 16, 14482 Potsdam, Germany
\\
\email{isabell.piantschitsch@uib.es}
}

\date{\today}

\abstract{The geometry of a coronal hole (CH) affects the density profile of the reflected part of an incoming global coronal wave (CW). In this study, we perform for the first time magnetohydrodynamic (MHD) simulations of fast-mode MHD waves interacting with CHs of different geometries, such as circular, elliptic, convex, and concave shapes. We analyse the influence these geometries have on the density profiles of the reflected waves and we generate the corresponding simulation-based time-distance plots. Within these time-distance plots we determine regions that exhibit specific density features, such as large reflected density amplitudes. In a further step, these interaction features can be compared to actual observed CW-CH interaction events which makes it possible to explain interaction parameters from the observed interaction events, such as the density structure of the reflected wave, which are usually difficult to comprehensively understand by only analysing the measurements. Moreover, we show that the interaction between a concave shaped CH and CWs, whose density profile include an enhanced as well as a depleted wave, can lead to reflected density amplitudes that are more than two times larger than the incoming ones. Another effect of the interplay between the constructive and destructive interference of the reflected wave parts is a strongly depleted region in the middle of the CW-CH interaction process. In addition, we show how important the choice of the path is that is used to generate the time-distance plots and how this choice affects the interpretation of the CW-CH interaction results.}


\maketitle

\section{Introduction}

Interactions between global coronal waves (CWs) and coronal holes (CHs) can be studied either by analysing the corresponding observations and measuring the involved parameters \citep{kienreichetal2013,olmedoetal2012,gopal2009,liu19,Chandra2022,Zhou2022,Mancuso2021}, or by performing numerical magnetohydrodynamic (MHD) simulations \citep{afanasyev2018,Piantschitsch2017,Piantschitsch2018a,Piantschitsch2018b,Piantschitsch2020} and thus trying to mimic the actual observed interaction events. Additionally, there is the possibility to directly compare the observational measurements with the numerical results and, therefore, to obtain an even more comprehensive insight in the whole interaction process. This study is the second one in a chain of papers, aiming to perform this direct comparison and to eventually reconstruct actual observed CW-CH interaction events. As a continuation of the first paper, in which the emphasis was on the initial density profile of the incoming wave \citep{Piantschitsch2023a}, the current study will focus on the geometry of the CH and its influence on the CW-CH interaction features.  

Why is it important to study these interaction features in the first place? The interaction between CWs and CHs results, among other effects, in the formation of reflected, refracted, and transmitted waves (collectively, secondary waves). These interaction features have been confirmed by observations \citep{Long2008, gopal2009,kienreichetal2013, liu19,olmedoetal2012, Veronig2006} as well as by numerical simulations \citep{Piantschitsch2017,afanasyev2018, Piantschitsch2018a,Piantschitsch2018b}. One reason for the relevance of studying these interactions is the still pending explanation for ambiguous phase speed measurements of secondary waves \citep{gopal2009,podladchikova2019}. Another motivation for this study is the fact that CW-CH interaction features can provide crucial information about the characteristics of CHs themselves, particularly about their boundaries and, hence, about the prediction of high-speed solar wind streams \citep{riley2015,Hofmeister2022}. In addition to that, recent studies showed that performing simulations of CW-CH interaction might be able to explain certain puzzling features in time-distance plots of actual observed interaction events \citep{Piantschitsch2023a}.

Which parameters are of importance in the interaction process? The two main protagonists in the interaction process are obviously CHs and CWs. CWs are defined as large-scale propagating disturbances in the corona and can be observed over the entire solar surface. Usually they have been denoted as "EIT waves" because they were directly observed for the first time by the Extreme-ultraviolet Imaging Telescope (EIT; \citealt{Delaboudiniere1995}) onboard the Solar and Heliospheric Observatory \citep{Domingo1995,thompson1998}. These propagating coronal disturbances are often also referred to as EUV waves or coronal bright fronts, and are commonly associated with energetic eruptions such as coronal mass ejections (CMEs) \citep[see e.g.][]{vrsnaklulic2000}.  The amplitude of a CW is a crucial factor when it comes to the analysis of CW-CH interaction. The amplitude is usually expressed as a compression factor that has a value of around $1.1$ or less for CWs with moderate speed \citep{Warmuth2015} (which will be used in the setup for this study) but can reach compression factors of up to $1.5$ which then are visible even in the lower chromosphere, and are called Moreton waves \citep{Moreton60}. This density amplitude parameter is derived from intensity measurements of CWs \citep{Muhr2011} together with the relation between intensity and density that can roughly be described as $\rho/\rho_0 \sim \sqrt{I/I_{0}}$ \citep[see][]{Warmuth2015}. 

CHs are the second main component in the interaction process and represent the focus in this study. CHs are low density, low temperature and weak magnetic field regions that are known as primary source of fast solar wind streams. Interacting with the ambient slow solar wind, these result in stream interaction regions and may cause geomagnetic effects \citep{Cranmer2009}. CH parameters, such as their area, boundary geometry, as well as longitudinal and latitudinal extent, affect predictions of high-speed solar wind streams \citep{Hofmeister2022,Samara2022,riley2015}. Among the on-disk CHs that have been studied so far, we can distinguish the ones at the poles from those located in the disk center \citep{Heinemann2020}. While polar CHs usually cover quite large areas, the CHs in the disk center, which tend to appear in more active periods of the solar cycle \citep{Cranmer2009}, are smaller in size. Some CHs at lower latitudes can exhibit long convex and concave boundaries as well, however, many of their boundaries are quite often similar to circular and elliptic shapes. 

Which parameters are available to describe CHs? Several parameters that characterise CHs have been measured and analysed increasingly detailed within the last decade. The most relevant parameter for our simulations is the density inside a CH, which is found to drop on average $30\%$ to $70\%$ with respect to the quiet Sun \citep{DelZanna1999,DelZanna2018,Saqri2020,Heinemann2021}. Another crucial parameter is the coronal hole boundary (CHB) width. It provides information whether there is a sharp density drop or a smooth gradient from the quiet Sun to the interior of the CH \citep{Heinemann2019}. The parameter which is at the centre of this study is the shape of the CHB. It plays an increasingly important role in recent studies. The detection of CHBs has become a crucial topic on its own and has led to the development of several different methods that aim to improve this detection, such as CATCH \citep{Heinemann2019} which is a threshold-based extraction method, and the Multi-Channel Coronal Hole Detection method developed by \citet{Jarolim2021} which is based on convolutional neural networks. Recently, these and other CHB detection methods have been compared to each other with regard to their observational uncertainty \citep{Linker2021,Reiss2021}.

How can we compare the simulation results to the observational measurements? The parameters involved in the observed interaction process are hard to derive unambiguously, as they are derived from intensity measurements and are based on the temperature dependent filtergrams in EUV. The analysis of the interaction parameters, such as the phase speed of the secondary waves, still mostly relies on studying the corresponding time-distance plots \citep[e.g.,][]{liu19,Chandra2022,Zhou2022,Mancuso2021}. In this study, we create time-distance plots based on the simulations to analyse the behaviour of the secondary waves in an analogous way as it is usually done in the observations. These results, combined with the results from \citet{Piantschitsch2023a}, will be used in a next step to partially reconstruct observational time-distance plots and, therefore, to directly compare the wave features from observations to those generated by simulations.

Until now, particularly the influence of the CH density and the initial density amplitude of the incoming wave have been studied in simulations of CW-CH interaction events \citep{afanasyev2018, Piantschitsch2017, Piantschitsch2018a, Piantschitsch2018b}. Other theoretical studies have taken into account the incident angle of the incoming wave and have provided analytical terms to estimate the density amplitude and the phase speed of reflected and transmitted waves of a CW-CH interaction \citep{Piantschitsch2020, Piantschitsch2021}. And very recently, the influence of a realistic density profile of the incoming wave, including an enhanced as well as a depleted wave part, has been studied for the first time \citep{Piantschitsch2023a}. However, the geometry of the CH, which is a crucial parameter in the interaction process, has not been included in any MHD simulation so far.

 The aim of this paper is to study the influence of the CH geometry on the interaction between CWs and CHs. The CH geometry and the resulting interaction effects will be analysed especially with regard to understanding and reconstructing observational time-distance plots. By generating the simulation-based time-distance plots and comparing them directly to those based on observations, we aim to provide a tool for deriving parameters from actual observed interaction events that are usually quite difficult to obtain directly from the measurements.

The paper will be structured in the following way:  Section \ref{section2} will be dedicated to the initial setup of the simulations and the description of the numerical scheme. In addition, some examples of CHs in observations will illustrate the motivation for the setup. In Section \ref{section3}, we present the simulation results of the CW interacting with the different CH geometries and we analyse the temporal evolution of the different corresponding 2D density structures. Section \ref{section4} is dedicated to the analysis of the simulation based time-distance plots and to identifying different representative areas of interest within these plots. We conclude in Section \ref{section5}.

\section{Numerical Setup}
\label{section2}

\subsection{Equations and algorithm}

We performed 2.5D simulations of fast-mode MHD waves, which exhibit a realistic initial density profile (enhanced and depleted wave part), that interact with low density regions of different geometries, representing various possible shapes of CHs. In our simulation code we use the MHD equations of continuity, momentum and induction, including the standard notations for the variables:
\begin{equation}
\frac{\partial{\color{black}\rho}}{\partial t}+\nabla\cdot(\rho \boldsymbol{v})=0,
\end{equation}
\begin{equation}
\frac{\partial(\rho \boldsymbol{v})}{\partial t}+\nabla\cdot\left(\rho \boldsymbol{vv}\right)-\boldsymbol{J}\times \boldsymbol{B}=0,
\end{equation}
\begin{equation}
\frac{\partial \boldsymbol{B}}{\partial t}-\nabla\times(\boldsymbol{v}\times \boldsymbol{B})=0.
\end{equation}

In this study, we consider an idealized case with a background magnetic field that is homogeneous and zero pressure all over the computational box. This is also the reason why no energy equation needs to be included in this setup.

The code we used to run the simulations has been applied in several former studies already \citep[e.g.,][]{Piantschitsch2017,Piantschitsch2018a,Piantschitsch2018b,Piantschitsch2020,Piantschitsch2023a}. The code is based on the so-called Total Variation Diminishing Lax-Friedrichs (TVDLF) method, which is a fully explicit scheme and was first described by \citet{toth1996}. To numerically solve the standard MHD equations (see Equations (1) - (3)) we make use of an alternating direction implicit method (ADI), which allows us to apply the TVDLF method in each coordinate direction one after each other. By including the Hancock predictor method \citep{VanLeer1984}, we obtain second-order temporal and spatial accuracy. The Hancock predictor step is based on the idea that within our computational grid we use first cell averages to predict the values of the conserved quantities at the cell edges at an auxiliary time step. Subsequently, the predicted values can be used to update the solution. A stable behaviour near discontinuities is guaranteed by applying the so-called Woodward limiter (for details see \citet{VanLeer1977} and \citet{toth1996}). To be able to also incorporate source terms in our scheme we included a four-stage Runge-Kutta method in the code. For numerical reasons, the Runge-Kutta operater is split up into two separate steps; the first step consists of one half of the time step and is
applied before the TVDLF-operater, the other step consists of the other half of the original Runge-Kutta time step but is applied after the TVDLF-operator.

We use Neumann boundary conditions at all four boundaries. In our simulation code this means that that the two rows of ghost cells at every boundary exhibit the same numerical value as the ones of the corresponding cell at the very edge of the actual computational grid. The size of the computational box is equal to 1.0 both in the $x-$ and the $y-$direction. We perform the simulations using a resolution of $300\times300$.

\subsection{Initial conditions}

For the CW simulation we apply two different initial conditions reflecting 1) a simple purely enhanced incoming wave pulse and 2) a more ``realistic'' profile. When we talk about a ``realistic'' density profile in this study, we refer to a CW profile that includes an enhanced and a depleted wave part, as it has been observed, e.g., in \citet{Muhr2011}. In our simulations we focus on CWs with an amplitude compression factor of around $1.1$ that are typical for CWs with moderate speeds \citep{Warmuth2015}.

The enhanced part of the initial incoming wave is excited in the following way:
\begin{equation}
     \rho(x) = 
    \begin{cases}
        \Delta\rho\cdot \cos^2(\pi\frac{x-x_0}{\Delta x})+\rho_0 & 0.4\leq x\leq0.6, \\
        \qquad \qquad 0.1   & \:\:\:\quad x\geq0.8, \\
        \qquad \qquad 1.0 & \:\:\qquad\text{else},
    \end{cases}
    \label{def_rho}
\end{equation}
\noindent where $\triangle\rho= 0.1$, $x_0=0.5$, $\triangle x=0.2$, and $\rho_0=1.0$. While for the rest of the variables
\begin{equation}
    v_x(x) = 
    \begin{cases}
        2\cdot \sqrt{\frac{\rho(x)}{\rho_0}} -2.0& \quad0.4\leq x\leq0.6, \\
        \:\:\qquad \qquad0 & \:\:\qquad\quad\text{else},
    \end{cases}
    \label{def_v}
\end{equation}
\begin{equation}
    B_z(x) = 
    \begin{cases}
        \:\:\rho(x) & \quad0.4\leq x\leq0.6, \\
        \:\: 1.0 & \:\:\qquad\quad\text{else},
    \end{cases}
    \label{def_B}
\end{equation}
\begin{equation}
B_{x}=B_{y}=0,\qquad\quad0\leq x\leq0.9,
\end{equation}
\begin{equation}
v_{y}=v_{z}=0,\qquad\quad0\leq x\leq0.9.
\end{equation}

The rear and depleted part of the incoming wave is excited in an analogous way, simply by choosing $\triangle\rho= -0.2$ and $x_0=0.2$. The range in which the density, the velocity and the magnetic field values are defined is shifted to $0\leq x\leq0.4$. This particular excitation generates a wave with the desired features propagating only toward the CH, that is, to the right in our coordinate system.

\begin{figure}[ht!]
\centering\includegraphics[width=\linewidth]{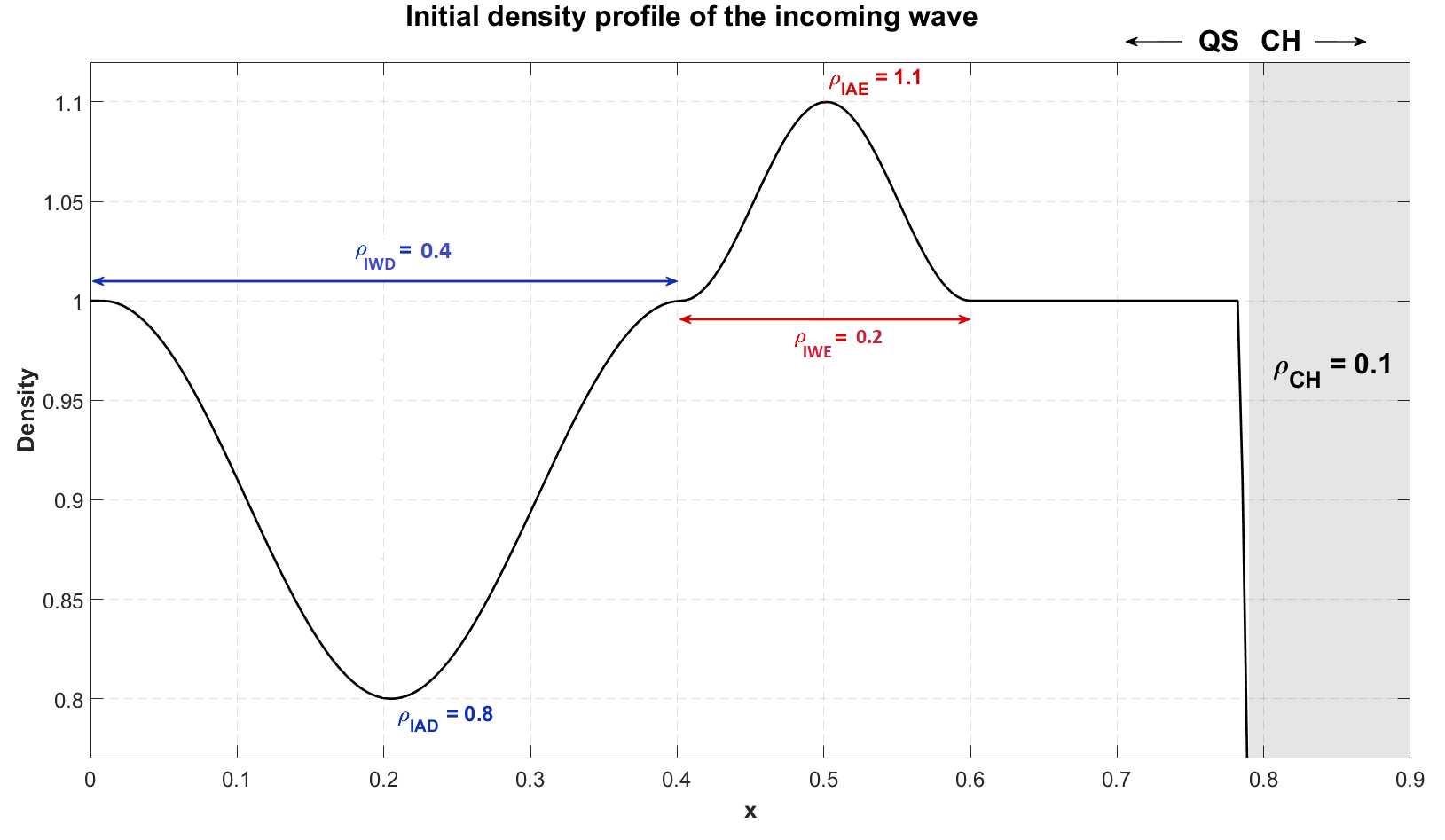}
\caption{Initial density profile of the incoming wave with a CH density of $\rho_{CH}=0.1$, a sharp density drop representing the CHB. We choose $\rho_{IAE} = 1.1$ for the initial enhanced amplitude (which corresponds to $\triangle\rho=0.1$, $\rho_0=1.0$, and $x_0=0.5$ in Equation \ref{def_rho}), $\rho_{IAD} = 0.8$ for the initial depleted amplitude (which corresponds to $\triangle\rho=-0.2$, $\rho_0=1.0$, and $x_0=0.2$ in Equation \ref{def_rho}), $\rho_{IWE} = 0.2$ for the initial width of the enhanced part (which corresponds to $\triangle x=0.2$ in Equation \ref{def_rho}), and $\rho_{IWD} = 0.4$ for the initial width of the depleted wave part (which corresponds to $\triangle x=0.4$ in Equation \ref{def_rho})}.\label{init_cond_inc_wave}
\end{figure}

The results in \cite{Piantschitsch2023a} show that the strongest and most visible interaction effects  are obtained by combining a sufficiently small CH density, an initial density profile for the incoming wave that includes an enhanced and a depleted part, and a sharp gradient representing the CHB width. That is the reason why we use the following parameters as initial setup: the CH density is set to the value of $\rho_{CH}=0.1$, the CHB is considered to be a step function (sharp gradient), and we use amplitude values of $\rho_{IAE} = 1.1$ for the initial enhanced amplitude and $\rho_{IAD} = 0.8$ for the depleted amplitude, we also define $\rho_{IWE} = 0.2$ to be the initial width of the enhanced part, and $\rho_{IWD} = 0.4$ to be the initial width of the depleted part. The incoming CW is invariant in the $y$-direction, that is, the CW is a superposition of plane waves with all their wavevectors parallel to the $x$-direction. In Figure \ref{init_cond_inc_wave} one can see the full density profile of the incoming wave, the CH density as well as the density amplitudes and the widths for the enhanced, and the depleted part of the initial incoming wave. 

\begin{figure}[ht!]
\centering\includegraphics[width=\linewidth]{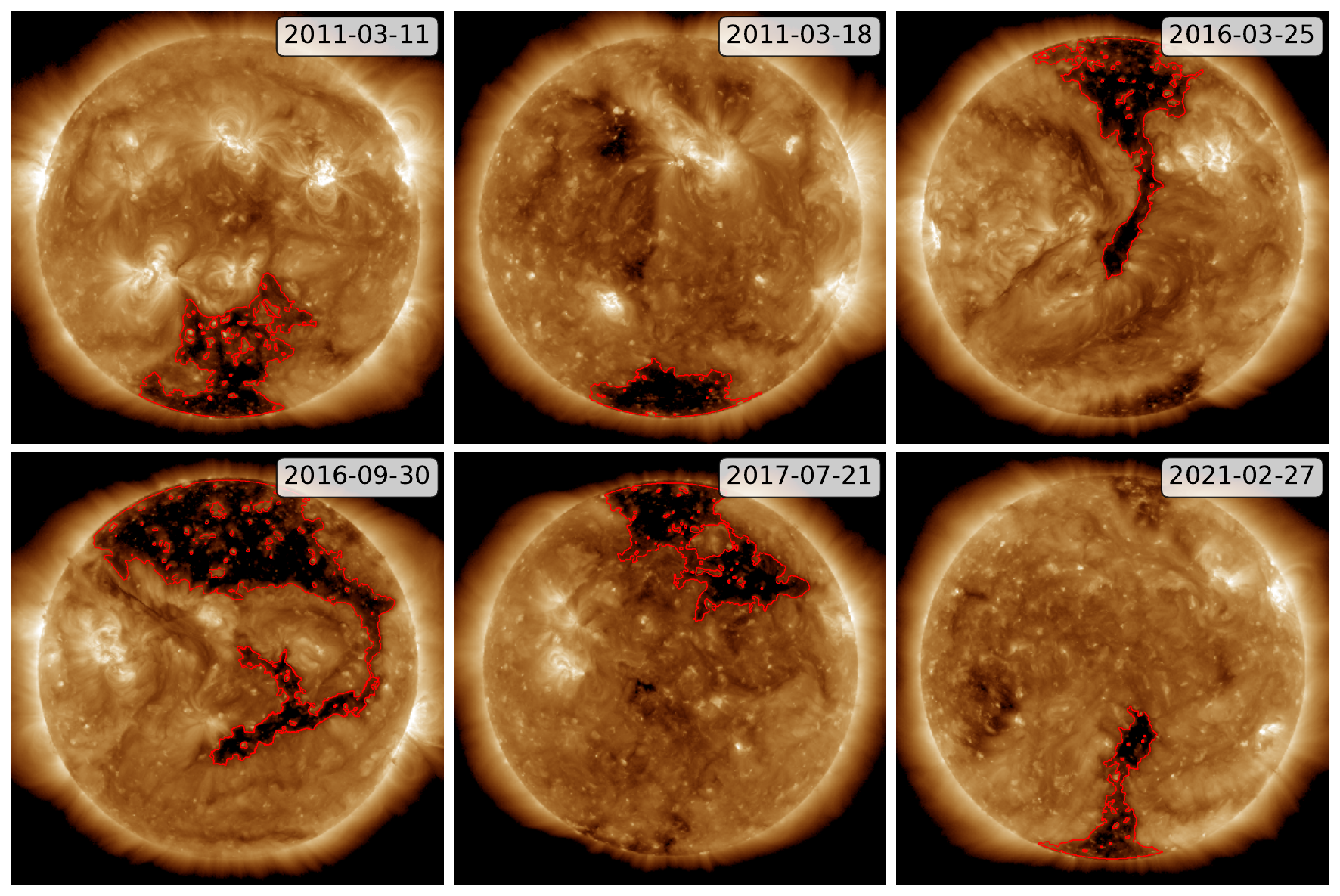}
\caption{Different polar CHs, which dominate the CH distribution in times of low solar activity, show longer convex and concave parts at their boundaries. CATCH \citep{Heinemann2019} was used to detect the boundaries of these CHs, whose outline is marked in each panel by red curves. The background images were obtained with the Atmospheric Imager Assembly \citep[AIA,][]{boerner2012,Lemen2012} on board the Solar Dynamics Observatory \citep[SDO,][]{Pesnell2012} in the 19.3 nm passband. }
\label{CH_shapes_observations_no2}
\end{figure}

\begin{figure}[ht!]
\centering\includegraphics[width=\linewidth]{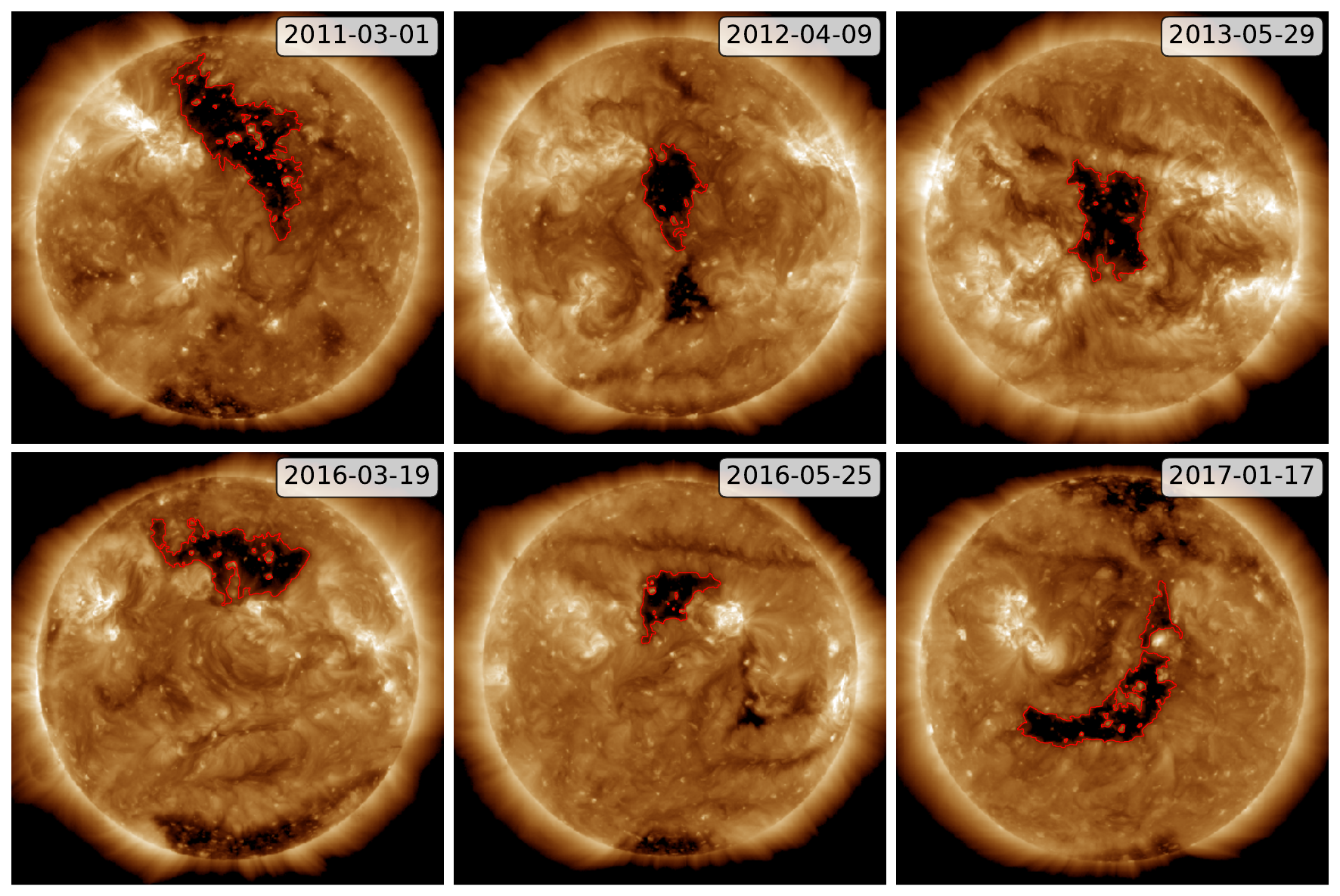}
\caption{Different CHs located in the disk center. These CHs, which tend to appear in more active periods of the Sun, often exhibit circular or elliptic shapes or a combination of these basic geometries. CATCH \citep{Heinemann2019} was used to detect the boundaries of these CHs, whose outline is marked in each panel by red curves. The background images were obtained with SDO/AIA in the 19.3 nm passband.}
\label{CH_shapes_observations_no1}
\end{figure}

In our simulations we focus on simple geometries, such as circular, elliptic, concave and convex structures. That means that the CH geometries in the numerical setup are simulated in an idealized way, but yet motivated by the shapes detected in the observations. Figure \ref{CH_shapes_observations_no2} shows different polar CHs which dominate the CH distribution in times of low solar activity \citep{Cranmer2009}. Figure \ref{CH_shapes_observations_no1}, on the other hand, shows CHs in the disk center that are usually a bit smaller in size than the polar CHs. The examples of CHs in Figure \ref{CH_shapes_observations_no2} and Figure \ref{CH_shapes_observations_no1}, which are supposed to serve as a motivation for the choice of the CH structure in our simulations, do not exhibit perfect and simple geometrical shapes of course. However, every complex geometrical structure can be considered as a combination of simpler shapes which explains our approach to first focus on idealised shapes before moving on to complex structures. 

\begin{figure}[ht!]
\centering\includegraphics[width=0.99\linewidth]{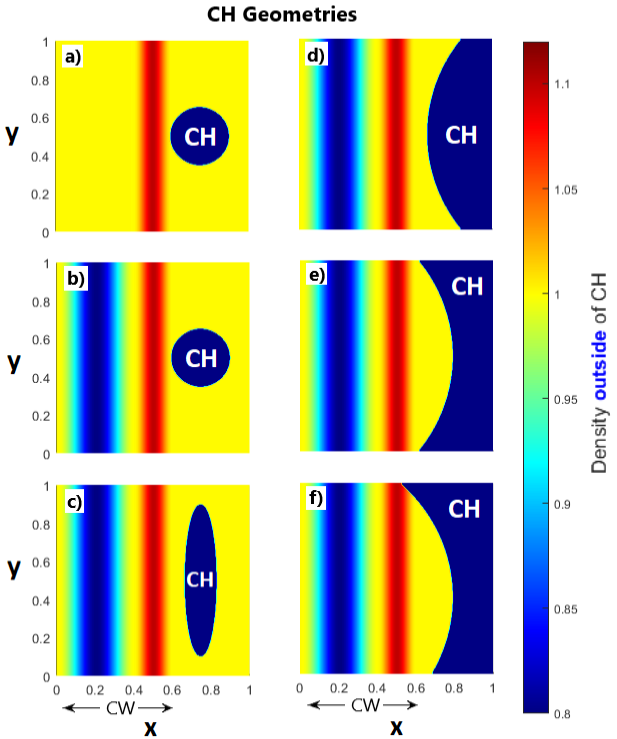}
\caption{Initial conditions for different CH geometries. Plot {\bf a)} shows an incoming wave that consists of a purely enhanced pulse (red) and a circular shaped CH. In plot {\bf b)} one can again see a circular shaped CH but this time with an incoming wave that is combining an enhanced pulse (red) with a depletion region at the rear wave part (blue). Plots {\bf c)} - {\bf f)} show incoming waves including enhanced as well as depleted parts and different CH shapes (elliptical in {\bf c)}, convex in {\bf d)}, concave symmetric in {\bf e)}, and concave asymmetric in {\bf f)}). Plots a), b), and c) refer to small CHs, which are commonly found in the disk center, whereas plots d), e), and f) refer to large CHs.}\label{init_cond_CH_shapes}
\end{figure}

Figure \ref{init_cond_CH_shapes} shows the different CH geometries we use in the initial setup of the simulations. We study cases of so-called closed shapes, such as small circular and elliptical shaped CHs, which are supposed to represent CHs that are more likely to be found in the disk center (see Figure \ref{init_cond_CH_shapes} a,b,c). The larger low density structures in Figure \ref{init_cond_CH_shapes} d, e, and f, mostly represent polar CHs. For the case of a large concave shape we consider the symmetric as well as the asymmetric situation (see Figure \ref{init_cond_CH_shapes}e and f). We also compare the results obtained by a simulation including a realistic initial density profile with a situation in which the incoming wave is a purely enhanced pulse (see Figure \ref{init_cond_CH_shapes}a). 

\begin{figure}[ht!]
\centering\includegraphics[width=0.94\linewidth]{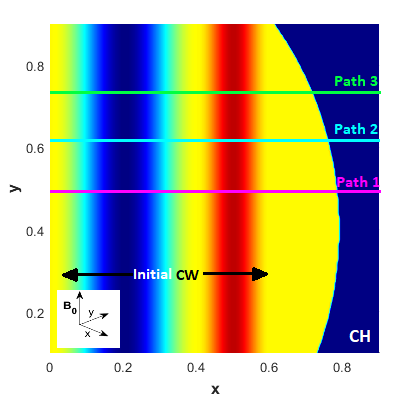}
\caption{Three different paths (Path1, Path2, and Path3) will be used along which the time-distance plots are generated. The dark blue vertical structure at the left denotes the depleted part of the initial incoming wave and the red vertical structure represents the enhanced part of the incoming wave (details of the amplitudes are shown in Figure \ref{init_cond_inc_wave}). The CH exhibits a non-symmetric concave density structure and has a CH density of $\rho_{CH}=0.1$. The background magnetic field is homogeneous and always pointing in the $z-$direction in this 2D setup.}\label{init_paths}
\end{figure}

The time-distance plots, which we use to analyse the properties of the incoming and the reflected wave, are created along three different paths within the computational box (see Path1, Path2, and Path3 in Figure \ref{init_paths}). In this way we try to address the fact that in observations one usually has to choose between several possible propagation slits that are used to generate observational time-distance plots \citep[see e.g., Figure 1 in][]{gopal2009}. Due to projection effects and the still rather low quality and accuracy of CW measurements, the right choice for the propagation direction in the observations can be quite challenging. We try to address this problem by choosing different paths along which the simulated time-distance plots are created and we compare the different results.

\section{Simulation results - 2D density structure}
\label{section3}

\subsection{Temporal evolution of the density structure}

In this section, the temporal evolution of a straight incoming wave (i.e., the propagation direction in every point of the incoming wave is the positive $x-$direction) interacting with CHs of different geometries is analysed, together with the corresponding time-distance plots and the density profiles of the incoming and the reflected wave.  

\begin{figure*}
\includegraphics[width=0.96\textwidth]{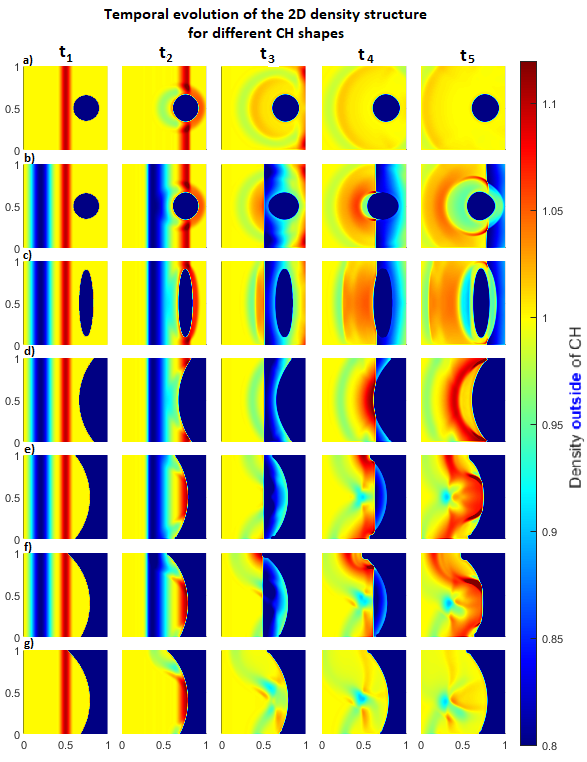}
\caption{Temporal evolution of the density distribution of a straight incoming wave interacting with CHs of different shapes. {\textbf{a)}} Interaction of a wave that consists of a purely enhanced pulse with a CH of circular shape. {\textbf{b)}} CH of circular shape and a CW that includes a density enhancement at the wave front and a depletion region at the rear part of the incoming wave \textbf{c)} elliptical shaped CH \textbf{d)} symmetric convex CH \textbf{e)} symmetric concave CH \textbf{f)} asymmetric concave CH \textbf{f)} purely enhanced pulse with an asymmetric concave CH}
\label{CH_shapes_comparison}
\end{figure*}

In Figure \ref{CH_shapes_comparison} one can see the temporal evolution of a CW interacting with CHs of different shapes (circle, ellipse, convex, concave symmetric, and concave asymmetric). In the case of a small circular shaped CH, we analyse two situations. First, a setup that includes a purely enhanced amplitude as an incoming wave (panel a) in Figure \ref{CH_shapes_comparison}), and second, a setup that considers a realistic initial wave density profile (enhanced and depleted wave part) for the interaction with the CH (panel b) in Figure \ref{CH_shapes_comparison}). This comparison between a purely enhanced and a realistic initial density profile is repeated in panels f) and g), but in this case including an asymmetric and concave shaped CH. These direct comparisons between two different initial incoming waves are supposed to show how significantly a realistic density profile changes the density distribution throughout the interaction process (for details see \citet{Piantschitsch2023a}). Panels b) - f), on the other hand, always use a realistic initial wave density profile but compare the effects the different CH shapes have on the density distribution of the CW-CH interaction. 

The results of the temporal evolution of the 2D density structure are given in Figure \ref{CH_shapes_comparison} with focus on various CH shapes. In all different cases, one can see how an incoming wave interacts with CHs of different shapes. The incoming wave gets reflected at the CHB and undergoes phase changes that eventually lead to superposition effects, that is, there is an interplay between the constructive and destructive interference of the reflected and the incoming wave parts \citep[cf.\,Figure 10 in][]{Piantschitsch2023a}. In the case of a realistic initial wave density profile (Figure \ref{CH_shapes_comparison}b) - f)) this is especially interesting because the interaction between the already reflected part of the enhanced incoming wave and the still incoming depleted part of the incoming wave lead to large depleted density areas and to a time period during which no wave part reaches values above background density (see $t_3$ in panels d) to f)). Subsequently, when the depleted part of the incoming wave starts interacting  with the CH and hence starts entering a phase changing process as well, the depleted amplitude of the reflected wave starts exhibiting larger density values again (still below background density though). Eventually, one part of the reflected wave gets pushed towards the background density level until it reaches amplitude values above background density level. These phase changes of the enhanced and the depleted incoming wave part, combined with the corresponding effects of constructive and destructive interference, that get intensified or reduced depending on the CH shape, lead to quite different density structures (see $t_5$ from a) to g)).

A comparison between panel a) and b), but also between f) and g), shows that an interaction including a realistic initial density profile leads to larger reflected density values (see red structures in $t_5$) and to an almost purely depleted density structure in the middle of the interaction process (see $t_3$) (for details regarding this comparison, see \citet{Piantschitsch2023a}). Depending on the CH geometry, the effects of large reflected density values and depleted areas are more or less pronounced. While the small and closed CHs (circular and elliptic) lead to quite small reflected density amplitudes, the large concave and convex CH structures exhibit much stronger interaction effects. The dark red structures at $t_5$ in panels e) and f) show that the reflected density amplitude can obtain values even larger than the initial incoming density amplitude and that the depleted area in the middle of the interaction process reaches values much below the initial incoming depleted wave part. As we will see later, when we analyse the density profiles in detail in Section \ref{section4}, the reflected density amplitude in the asymmetric concave case is able to reach values even two times larger than the values of the incoming density amplitude. In Table \ref{table} one can see the minimum and the maximum density values within the reflected wave for the different CH shapes seen in Figure \ref{CH_shapes_comparison}. We can see that the largest but also the smallest density value during the interaction process is obtained by the asymmetric concave shaped CH ($\rho_{max}\approx1.24$ and $\rho_{min}\approx0.73$), whereas the least intense interaction effects are a result of the interaction with a small circular shaped CH ($\rho_{max}\approx1.06$ and $\rho_{min}\approx0.78$). As already mentioned above, all these different density values occur due to the interplay between the constructive and destructive interference of the different reflected and incoming wave parts.

\begin{table}[ht]
\centering
\begin{tabular}{|c|c|c|}
\hline
\cellcolor[HTML]{DAE8FC}\textbf{CH shape} & \begin{tabular}[c]{@{}c@{}}\cellcolor[HTML]{F5A5A0}{\color[HTML]{000000}$\rho_{max}$ of} \\ \cellcolor[HTML]{F5A5A0}{\color[HTML]{000000}reflection}\end{tabular} & \begin{tabular}[c]{@{}c@{}} \cellcolor[HTML]{89B7F8}{\color[HTML]{000000}$\rho_{min}$ of}\\ \cellcolor[HTML]{89B7F8}{\color[HTML]{000000}reflection}\end{tabular} \\ \hline
circle           &   $\approx 1.06$                                                              &   $\approx 0.78$                                                                \\ \hline
ellipse           &       $\approx 1.07$                                                             &        $\approx 0.76$                                                           \\ \hline
convex            &   $\approx 1.13$                                                                    & $\approx 0.74$                                           \\ \hline
concave symm.     &   $\approx 1.2$                                                    &  $\approx 0.74$                                           \\ \hline
concave asymm.    &  $\approx 1.24$                                                                  &      $\approx 0.73$                                                            \\ \hline
\end{tabular}

\vspace{10px}
\caption{Minimum and maximum density values of the reflected wave after the interaction with different CH shapes (circle, ellipse, convex, concave symmetric, and concave asymmetric), considering and initial incoming wave profile with $\rho_{IAE} = 1.1$ for the initial enhanced amplitude, $\rho_{IAD} = 0.8$ for the depleted amplitude, $\rho_{IWE} = 0.2$ for the initial width of the enhanced part, and $\rho_{IWD} = 0.4$ for the initial width of the depleted part. }

\label{table}
\end{table}

\begin{table*}[ht!]
\centering

\includegraphics[width=0.8\textwidth]{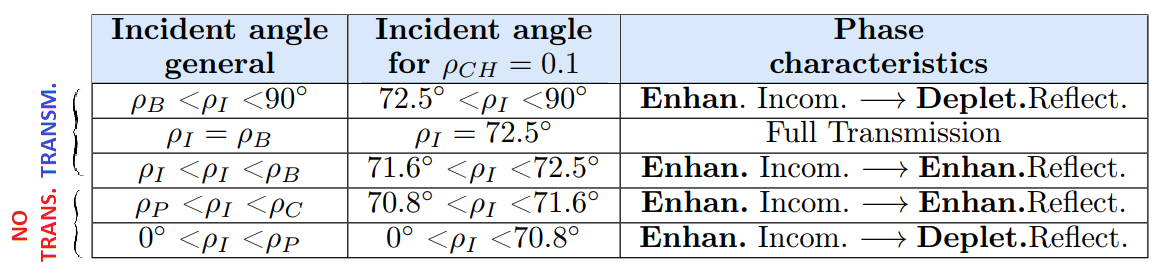}

\vspace{10px}
\caption{Representative angles, transmission features and phase characteristics for general incident angles and for the case of $\rho_{CH}=0.1$. For a CH density of $\rho_{CH}=0.1$ the Brewster angle, $\theta_B$, is equal to $72.5^\circ$, the Phase inversion angle, $\theta_P$, is equal to $70.8^\circ$, and the Critical angle, $\theta_C$, is equal to $71.6^\circ$. One can see that incident angles above the Critical angle lead to transmitted waves whereas incident angles below the Critical angle result in pure reflection. The Brewster angle $\theta_B$ represents a situation of perfect transmission, and the Phase inversion angle $\theta_P$ separates enhanced from depleted reflection. For linear waves this implies that an enhanced incoming wave with an incident angle between the Brewster angle and the Phase inversion angle gets reflected as an enhanced wave, whereas incoming waves that exhibit an incident angle outside of this range undergo a phase change and propagate as depleted reflections.}
\label{table2}
\end{table*}

In \citet{Piantschitsch2023a} it has been shown that large reflected density amplitudes can be obtained by combining a sufficiently small CH density with a small incident angle and a realistic initial wave density profile, including an enhanced and a depleted part. (Here, the incident angle is defined as the angle that the wavevector forms with the CHB.) If only one of these three conditions is not satisfied, a large reflected density amplitude cannot be reached. This result gets confirmed again in Figure \ref{CH_shapes_comparison}f where we can see that a concave shaped CH with a small CH density leads to quite strong interaction effects. However, these large density values are not obtained in panel g) in which the incoming wave is considered to be a purely enhanced pulse. In other words, a purely enhanced incoming wave is sufficient to significantly reduce the strong interaction effects seen in panel f). This shows once again the importance of the right choice for the initial wave profile in MHD simulations including CWs. 

\subsection{Representative incident angles}

The question now is how to explain the different large and small density values within the reflected part of the wave. In other words, do we know more details about the effects of constructive and destructive interference of the reflected wave parts in the different cases of CH shapes? Already in \citet{Piantschitsch2021} we provided analytical expressions for three representative incident angles that give us information about the phase of the reflected wave and also about whether there is a transmission through the CH happening. All of these angles depend only on the density contrast $\rho_c$ (ratio of the CH density to the density of the quiet Sun) which is equal to CH density in our case because we consider the background density to be equal to $1.0$. The first angle in this list of representative incident angles is the so-called Critical angle, $\theta_C$,

\begin{equation}\label{thetc}
    \theta_{\rm C}={\rm cos^{-1}} \left(\sqrt{\rho_{\rm c}}\right).
\end{equation}

This angle separates the situation of a full reflection from a situation that includes transmission, that is, incident angles above the Critical angle lead to transmitted waves whereas incident angles below the Critical angle result in pure reflections (for details see Figure 7 and Figure 8 in \citet{Piantschitsch2021}).

The second representative angle is the so-called Brewster angle, $\theta_B$, 

\begin{equation}\label{thetb}
    \theta_{\rm B}={\rm cos^{-1}} \left(\sqrt{\frac{\rho_{\rm c}}{1+\rho_{\rm c}}}\right),
\end{equation}
and it represents a situation in which perfect transmission takes place, that is, there is no reflection at the CHB at all. The third important incident angle which separates enhanced from depleted reflection is the  so-called Phase inversion angle, $\theta_P$,

\begin{equation}\label{thetS}
    \theta_{\rm P}={\rm cos^{-1}} \left(\sqrt{\frac{\rho_{\rm c}+\rho^2_{\rm c}}{1+\rho^2_{\rm c}}}\right),
\end{equation}
where $\rho_{\rm c}$ denotes again the density contrast. For linear waves this implies that an enhanced incoming wave with an incident angle between the Brewster angle and the Phase inversion angle gets reflected as an enhanced wave, whereas incoming waves that exhibit an incident angle outside of this range undergo a phase change and propagate as depleted reflections. For a CH density of $\rho_{CH}=0.1$, for instance, we know that the Brewster angle, $\theta_B$, is equal to $72.5^\circ$, the Phase inversion angle, $\theta_P$, is equal to $70.8^\circ$, and the Critical angle, $\theta_C$, is equal to $71.6^\circ$. One can see that in the case of a small CH density these values are very close to each other. That means that already a small change in the incident
angle is sufficient to turn a case of full transmission into
a case of no transmission at all. This result provides a crucial information for the interpretation of CW–CH interaction effects in observational data. Table \ref{table2} shows these transmission features and the phase characteristics for a general incident angle and provides values for the three different representative angles in the case of $\rho_{CH}=0.1$ (for more details also see \citet{Piantschitsch2021}). We want to emphasize that the analytical expressions for the representative angles and the corresponding phase characteristics were derived using linear wave theory. However, since in this study we are interested in linear and weakly non-linear MHD waves (up to a compression factor of $1.1$), which are also motivated by observational measurements of CWs \citep{Muhr2011,Warmuth2015}, these theoretical results seem to be a sufficiently reasonable basis for the interpretation of the simulation results.

A situation that considers an interaction with a concave shaped CH includes in fact many different incident angles and, hence, leads to a superposition of reflected waves exhibiting different phases and different amplitudes.
When we look again at Table \ref{table2}, we can see that probably all four different incident angles are involved in the case of a concave shaped CH. This means that some incoming enhanced wave parts might get reflected as depletions, some might get reflected as enhanced waves, some might get fully reflected and some might travel through the CH as well (and analogously for the depleted incoming wave part). Moreover, the representative angles always depend on the CH density, and therefore, the density structure of an interaction with a concave shaped CH can look very different depending on the CH density. These considerations show how complex the interaction process can be, depending on the CH geometry, the CH density, and the initial incoming density profile.

\begin{figure*}[ht!]
\includegraphics[width=\textwidth]{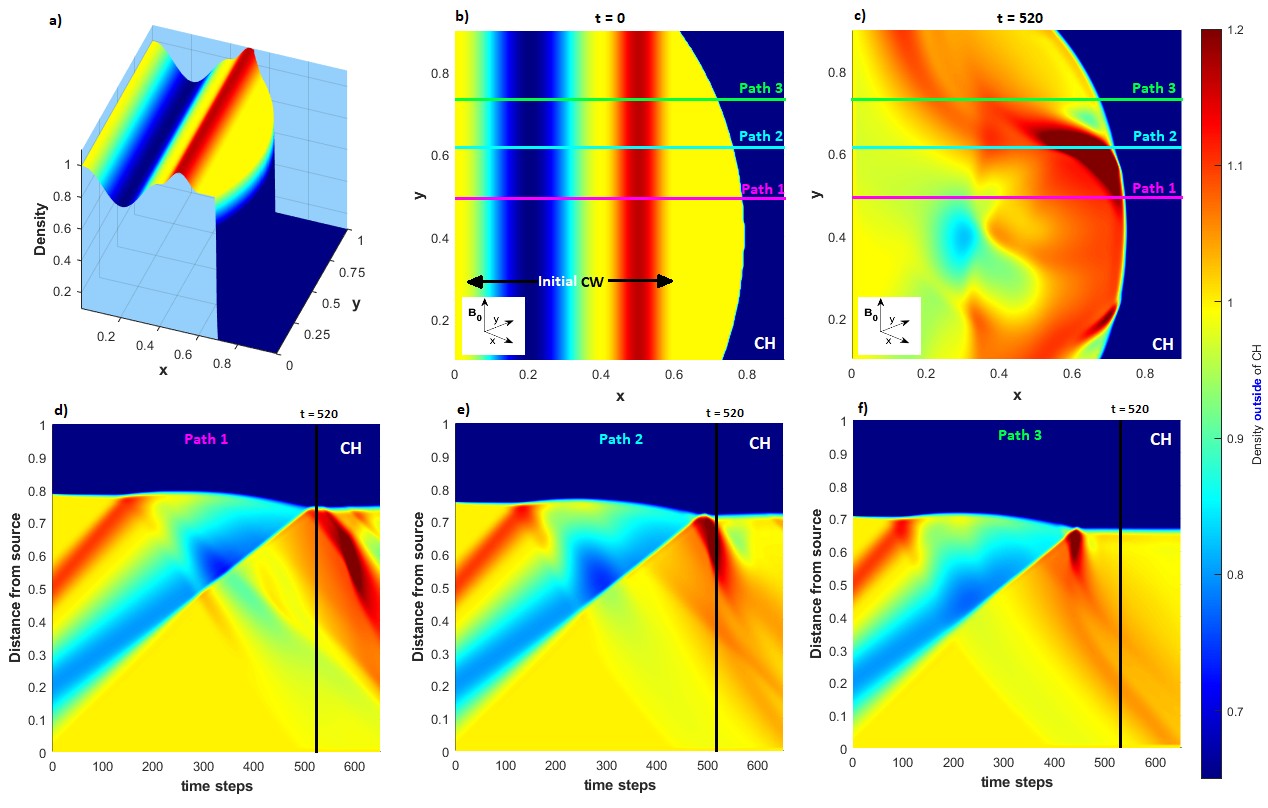}
\caption{\textbf{a)} This figure shows a 3D visualisation of the initial conditions of a straight incoming wave (including a density enhancement, red, and a depleted wave part, blue) interacting with an asymmetric concave area of low density which denotes the CH. \textbf{b)} One can see the 2D initial conditions at $t=0$, as in plot a). The magenta, turquoise and green lines denote the paths along which the different time-distance plots are generated. \textbf{c)} This plot shows the density distribution at the end of the temporal evolution ($t=650$) of the CW-CH interaction together with the three paths along which the time-distance plots are created. \textbf{d)-f)} These time-distance plots show the temporal evolution of the incoming and the reflected wave along Path 1 (magenta), Path 2 (turquoise), and Path 3 (green). The blue and green areas denote regions in which the density is below the background density, whereas the red and orange areas denote regions in which the density is larger than the background density. }
\label{init_cond_concave}
\end{figure*}

\section{Simulation results - Time-distance plots}
\label{section4}

\subsection{Time-distance plots depending on CH geometry}

The temporal evolution of the incoming and the reflected density profiles can be studied in more detail by analysing the corresponding line profiles and time-distance plots. In Figure \ref{init_cond_concave} we show the three different paths (at $t=0$ and $t=520$) along which the line profiles and the time-distance plots are created. We decided to choose different paths for analysing the density profiles due to the fact that in the observations we often have to face the difficulty of choosing a clear and unambiguous slit to generate time-distance plots. The propagation direction in the observations is usually not sufficiently clear and unequivocal to determine a unique propagation direction. This is the reason for us to use different paths for generating the line profiles in the MHD simulations, and therefore, for capturing the different propagation directions in the observations. 

\begin{figure*}[ht!]
\includegraphics[width=0.99\textwidth]{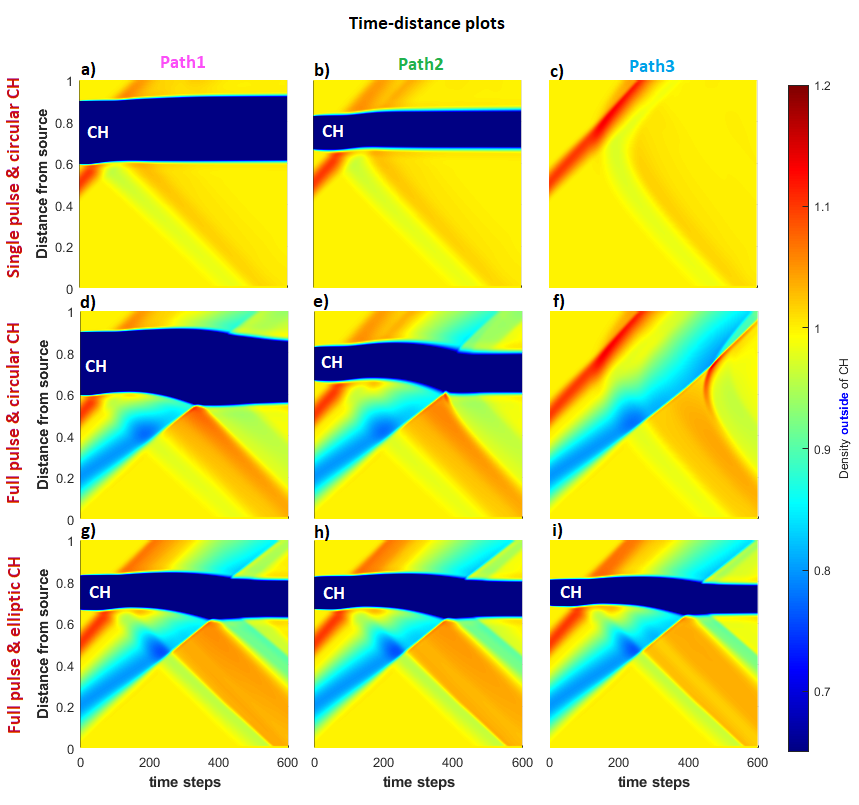}
\caption{Time-distance plots along Path1 (magenta), Path2 (green), and Path3 (blue) for different density profiles of the incoming wave and different CH geometries. The blue and green coloured regions denote areas in which the density is smaller than the background density (depletion) whereas the red and orange coloured regions denote areas in which the density is larger than the background density (enhancement). The dark blue horizontal regions denote the CH area. \textbf{a) - c):} Time-distance plots of an incoming wave that consists only of an enhancement, WITHOUT a depletion at the rear part of the wave interacting with a circular shaped CH. \textbf{d) - f):} Time-distance plots of an incoming wave with a density profile consisting of an enhancement at the wave front and a depletion area at the rear part of the wave, interacting with a circular shaped CH. \textbf{g) - i):} Time-distance plots of a full density profile (enhancement + depletion) interacting with an elliptic shaped CH and a CH density of $\rho_{CH}=0.1$.}
\label{time_dist_circle_ellip}
\end{figure*}

\begin{figure*}[ht!]
\includegraphics[width=0.99\textwidth]{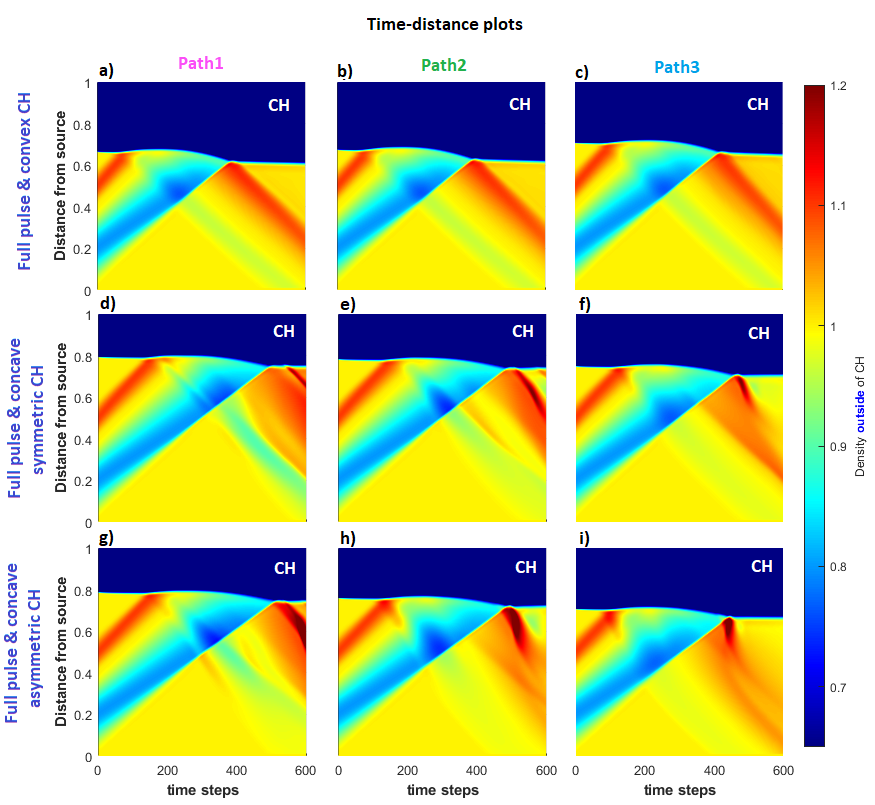}
\caption{Time-distance plots along Path1 (magenta), Path2 (green), and Path3 (blue) for different density profiles of the incoming wave and different CH geometries. The blue and green coloured regions denote areas in which the density is smaller than the background density (depletion) whereas the red and orange coloured regions denote areas in which the density is larger than the background density (enhancement). The dark blue horizontal regions denote the CH area. All incoming waves consist  of an enhancement at the wave front and a depletion area at the rear part of the wave. The CH density for all cases is $\rho_{CH}=0.1$. \textbf{a) - c):} Interaction with a convex shaped CH. \textbf{d) - f):} Interaction with a concave symmetric CH. \textbf{g) - i):} Interaction with a concave asymmetric CH.}
\label{time_dist_convex_concave}
\end{figure*}

In Figure \ref{init_cond_concave}a one can see a 3D visualisation of the initial setup including the initial incoming wave with a realistic density profile and an asymmetric concave area of low density which denotes the CH. Plot b) shows the initial 2D density distribution of the simulation setup ($t=0$). In plot c) one can see the density distribution at $t=520$. Plots d) - f) of Figure \ref{init_cond_concave} show the time-distance plots that were created along the three different paths, Path1 (magenta), Path2 (turquoise), and Path3 (green). The blue and green areas denote regions in which the density is below the background density, whereas the red and orange areas denote regions in which the density is larger than the background density. The dark blue regions denote the CH. The vertical black lines in \ref{init_cond_concave}d), e), and f) refer to the same density profiles as the corresponding paths in \ref{init_cond_concave}c). In this plot, it is already obvious how the density structure in the time-distance plots varies depending on the path we use to create it, that is, the time-distance plot corresponding to Path 1 shows stronger interaction effects than the time-distance plot based on Path 3.

\begin{figure*}[ht!]
\includegraphics[width=0.99\textwidth]{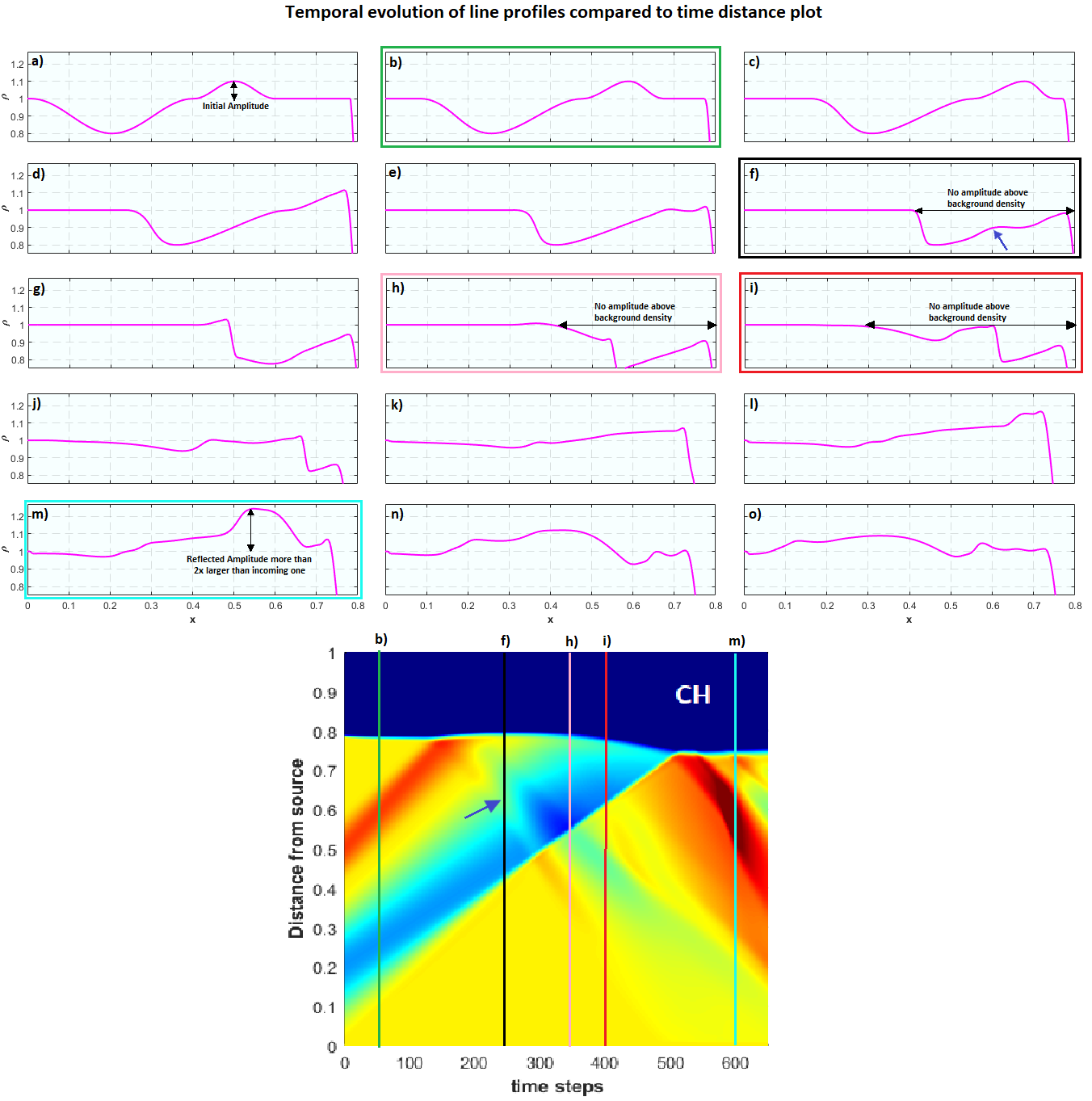}
\caption{Temporal evolution of the density wave profile (incoming and reflected wave) along Path1 (magenta) compared to the related time-distance plot of this interaction with a concave asymmetric CH. In \textbf{a)} one can see the initial density profile with an enhanced amplitude of $\rho_{Ampl}=1.1$ and a CH density of $\rho_{CH}=0.1$. Plot \textbf{b)} is also represented as vertical green line in the time-distance plot at the bottom and shows how the incoming wave still moves towards the CH. In \textbf{b)} one can see that no part of the density profile reaches a value above the background density $\rho_0=1.0$. A similar situation can be found in \textbf{h)} and \textbf{i)}. These profiles are denoted in the time-distance plot as black, pink and red vertical lines. The temporal evolution between \textbf{e)} and \textbf{j)} shows clearly how the first reflected wave part propagates through the depleted part of the incoming wave. In plot \textbf{m)} one can see that the amplitude of the reflected wave is able to reach a value that is more than two times larger than the density amplitude of the incoming wave. This large amplitude can also been seen as dark red structure in the time-distance plot below.}
\label{line_profile_time_dist}
\end{figure*}

Figures \ref{time_dist_circle_ellip} and \ref{time_dist_convex_concave} show the time-distance plots which correspond to the temporal evolution of the density distribution in Figure \ref{CH_shapes_comparison}. In Figure \ref{time_dist_circle_ellip} one can see the time-distance plots for the circular and the elliptical CH shape. Plots a)-c) show the situation of a purely enhanced incoming wave in combination with a circularly shaped CH. Here, the enhanced reflected part (dark orange structure pointing into the lower right corner of the plot) is barely visible as one could already see in density structure of Figure \ref{CH_shapes_comparison}a. Plots d)-f) depict the time-distance plots of the interaction between an incoming wave including a realistic initial density profile and a circular shaped CH, and plots g)-i) show such an interaction with an elliptically shaped CH. As we could already see in Figure \ref{CH_shapes_comparison}b and c, there is a region around time step $t=200$ in which we only detect density values below the background density (blue and light green areas). Such regions are not visible in the situation of the purely enhanced incoming wave in \ref{CH_shapes_comparison}a. Another result that was also already visible in the temporal evolution of Figure \ref{CH_shapes_comparison} is the fact that along all three different paths (for the realistic initial density profile), the width of the enhanced reflected wave is broader than the incoming one. However, in none of the situations the (enhanced) reflected amplitude is larger than the incoming one. The larger width of the reflected enhanced wave part can be explained by the large width of the initial depleted part of the incoming wave that. The density amplitudes of the reflected wave, which are smaller than the ones of the incoming wave in this case, can be explained by the shapes of the CHs which do not lead to sufficiently strong effects of constructive interference of the different reflected wave parts. Within the density structure of plots d)-i) one can also see a small bump which points into the direction towards the upper left part of the plots. This bump, which continues into the direction of the green and yellow area, depicts the motion of the first reflected wave part that travels through the still incoming rear part of the wave until it reaches (due to destructive interference effects) the smallest interaction density values. It finally propagates as a slightly depleted reflection (light green structure below the enhanced reflection) away from the CH.

Figure \ref{time_dist_convex_concave} shows the time-distance plots for a convex, a symmetrical concave, and an asymmetrical concave shaped CH. Similar to the cases of a small circularly and elliptically shaped CH, the interaction with a convex CH does not result in a large enhanced reflected amplitude with respect to the incoming one. The situation, however, for a concave shaped CH is different, as we can see in plots d)-i) of Figure \ref{time_dist_convex_concave}. The time-distance plots for the symmetric and the asymmetric concave shaped CH show a dark red area within the enhanced reflected wave part. Here, the reflected amplitude is more than two times larger than the incoming one. Moreover, one can see, that especially in the asymmetric concave case (g)-i)), the angle of the reflected wave is not the same as the incoming one, it is more bent towards the center of the time-distance plot. The large reflected amplitude structure (dark red structure) is caused by constructive interference effects when the incoming wave interacts with a concave shape. Generally, due to these effects of constructive and destructive interference, the density structure of the reflected wave appears to be much more complex in the concave case than in all other cases. Besides the non-uniform density structure of the enhanced reflected part, also the depleted part, which moves ahead within the reflected wave, shows a different behaviour than in the other cases (see d) - i)). Both parts, the enhanced as well as the depleted reflection, do not exhibit a uniform width or amplitude while moving away from the CH. This makes it challenging to determine a clear propagation direction in the simulations as well as in the observations.

In observations, the reflected phase speed is usually derived by drawing a line along the presumed direction of the reflected wave in the time-distance plots (in our simulations this is the dark red structure). And if the angle formed by the direction of the incoming wave and a vertical line is larger than the one formed by the direction of the reflected wave and a vertical line (see, e.g., Figure \ref{init_cond_concave}), we interpret the reflected phase speed to be larger than the incoming one. But this method is only valid if the actual incoming propagation direction is the same as the reflected one, which is usually not the case. That means that if we use the same path for generating the time-distance plots of the incoming AND the reflected wave, we have to be very careful. In the concave case, the dark red region in Figure \ref{CH_shapes_comparison}f (which is the large enhanced amplitude of the reflected wave) moves across the computational box. However, it does not only move in the negative $x-$direction but even points into the left upper corner of the computational box. That is the reason why we have a bent dark red structure in the corresponding time-distance plots. That also means that in this case it is dangerous to try to obtain the reflected phase speed by drawing a straight line, because this does not reflect the actual situation of the interaction process. The bent structure in the time-distance plots provides more information about a change of propagation direction of the reflected wave than about the actual phase speed. However, we can rely on the value of the large reflected amplitude. This result also shows the importance of a correct path choice for creating time-distance plots.

\begin{figure*}[ht!]
 \centering\includegraphics[width=0.7\textwidth]{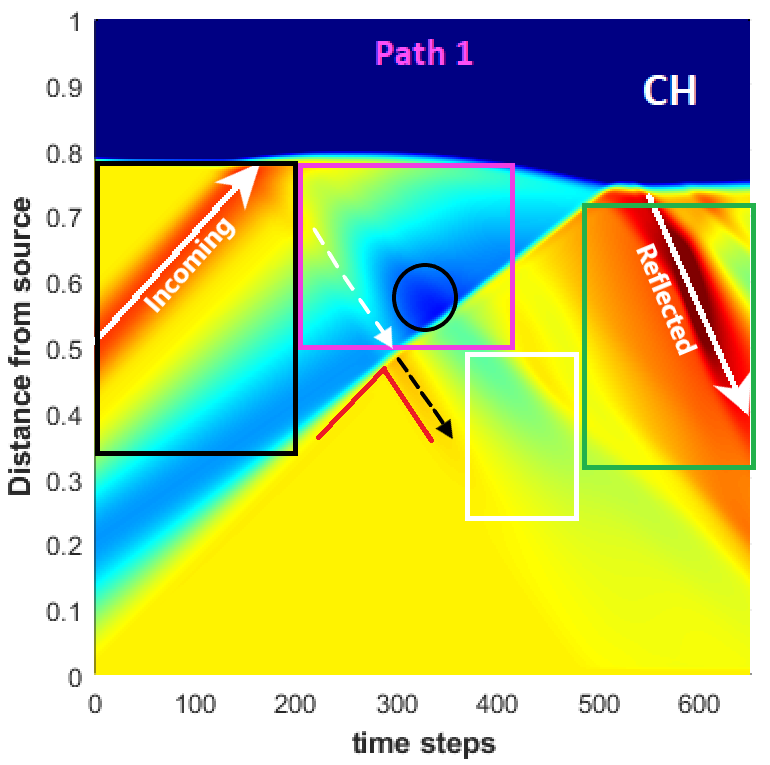}
\caption{Time-distance plot of the interaction between an incoming wave including a realistic density profile (enhancement + depletion) and an asymmetric concave shaped CH that has a CH density of $\rho_{CH}=0.1$ and an initial density amplitude of $\rho_{IAP}=1.1$, created along the central path (Path1, magenta). The black rectangle denotes the area in which the incoming wave is moving towards the CH and no interaction has taken place yet. The magenta rectangle shows the part of the interaction at which no density value above background density level can be detected and the black circle within the magenta rectangle denotes the area in which the density structure reaches its minimum value. The area that is enclosed by the green rectangle represents the region in which the reflected wave can be seen as an enhanced wave again, always reaching values above background density and even reaching density values more than two times the ones of the incoming wave (dark red area). The white rectangle denotes the area where the reflected wave travels mostly as a depletion and none of the strong enhanced reflected wave parts if visible yet. The white and the black dashed lines denote the paths of a first reflection that travels through the rear part of the still incoming wave.}
\label{time_dist_rectangular_regions}
\end{figure*}

In Figure \ref{line_profile_time_dist} we analyse in more detail the temporal evolution of the density profile for the central magenta path in case of the asymmetric concave shaped CH. In a) one can see the initial density profile, in b) the incoming wave propagates towards the CH which can also be seen along the vertical green line in the time-distance plot at the bottom. Plot c) shows how the incoming wave propagates towards the CH and is about to interact with the CHB. In plot d) the interaction has just begun and a first part of the wave enters the CH while another part of the wave front gets reflected at the CHB, undergoing a phase change. In plot f) of Figure \ref{line_profile_time_dist}, the front part of the incoming wave already got reflected (see blue arrows in f) and in the time-distance plot at the bottom) and interacts with the still incoming rear (depleted) part of the initial wave. This also explains the fact that in f) no density value above background density can be seen. The same is true for the line profiles in h) and i) where the low density values are also visible as blue and green structures in the time-distance plot. Plot l) already shows reflected density values that are clearly much larger than the initial density amplitude of the incoming wave. In m) one can see that after the entire incoming wave has interacted with the CHB, the amplitude of the reflection is able to reach a value more than two times larger than the initial enhanced amplitude of the incoming wave. This large amplitude can also be seen in the time-distance plot at the bottom of the figure, denoted along the turquoise vertical line.  In plot h) and in the time-distance plot below where the pink vertical line is located, one can see that the density amplitude reaches a value below the initial depleted density amplitude of the incoming wave. This is the result of the destructive interference between the depleted part of the reflection and the still incoming depleted wave part.

\subsection{Representative interaction features in the time-distance plots}
The final goal we aim to achieve with the help of the parameter studies in this paper is to compare the simulation results to actual observed interaction events. This motivates us to already determine specific regions within the simulation based time-distance plots which exhibit characteristic density features that can then be directly compared to observational time-distance plots. With that we try to derive interaction parameters from the observed interaction events which usually cannot be obtained directly from the measurements. In Figure \ref{time_dist_rectangular_regions} one can see how we divided the density structure of the time-distance plot into different regions of interest. The black rectangle denotes the part of the interaction at which the entire incoming wave still moves towards the CH until shortly before the front part of the wave starts interacting with the CHB. The red inclined structure within that black rectangle denotes the enhanced part of the incoming wave whereas the blue and green area denote the depleted part which follows the wave front during the propagation towards the CH. The area enclosed by the magenta rectangle shows the density structure in the time period during which no density value above background density level can be detected. Within this rectangle we see furthermore some reflected wave part that tries to push the rear part of the incoming wave backwards (white dashed line). This is the reflected (depleted) part of the enhanced incoming wave that is moving through the depleted still incoming wave part in the negative $x-$direction. After the first reflected part has completed travelling through the still incoming depletion it continues travelling in the negative $x-$direction and can now be seen as a light orange area (see also black dashed line). The black circle within the magenta rectangle denotes the area in which the density structure reaches its minimum value, it is located at the position where the smallest density value of the first reflected part of the incoming wave meets the smallest density values of the still incoming depletion. Hence, the interference leads to the minimum density value of the entire interaction process. The region enclosed by the white rectangle denotes the area where the reflected wave travels mostly as a depletion in the negative $x-$direction and none of the strongly enhanced reflected wave parts are visible yet. The area that is enclosed by the green rectangle represents the region in which the reflected wave can be seen as an enhanced wave again. In this area the wave reaches everywhere values above background density and in some parts it even reaches density values of more than two times the ones of the incoming wave (dark red area). This large density region starts appearing at the time when the rear part of the incoming wave has completed entering the CH and it is the result of the superposition of several different wave components propagating towards the central part of the computational box. We have to keep in mind that this interference is only possible due to the concave geometry of the CH, in all the other cases the reflected waves cannot reach larger density values than the incoming ones. In Figure \ref{time_dist_rectangular_regions} we also defined two directions for the incoming and the reflected wave (white arrows), the way they are usually defined in the observational time-distance plots. The simulation results show us, however, that one has to be careful when using time-distance plots to analyse incoming and reflected waves. Using only one slit may lead to wrong conclusions regarding the reflected phase speed. Another feature that is clearly visible in this figure is the deformation of the CH boundary due to the incoming wave. Until now, this deformation has not been investigated in detail from the observational point of view. However, it will be included in the direct comparison to an actual observed event in our follow-up study.

\section{Conclusions}
\label{section5}

In this study, we perform for the first time MHD simulations of CWs interacting with CHs that exhibit different geometries. The initial simulation setup is based on the results of \citet{Piantschitsch2023a} where we analysed the effects of a CW density profile, that includes an enhanced and a depleted wave part, on the CW-CH interaction process. In this paper, we focus on the shape of the CHs by considering different basic geometries, such as circular, elliptical, convex and concave density structures. We analyse the density profiles of the incoming and the reflected waves as well as the temporal evolution of the 2D density structure during the whole interaction process. Moreover, we generate the corresponding time-distance plots, and within these plots, we determine specific regions that exhibit characteristic density features which can, in a further step, be compared to actual observed CW-CH interaction events. 

The main results are summarised as follows:

    \begin{enumerate}
    \item For the first time, simulations of the interaction between CHs of various shapes and CWs that include an enhanced as well as a depleted part were performed. Depending on the shape of the CH, differences in the interaction effects, such as large reflected density amplitudes and entirely depleted areas within the density structure, could be detected (see Figure \ref{CH_shapes_comparison}).

    \vspace{0.3cm}
    \item Among the circular, elliptic, convex and concave CH geometries that we studied in the simulations, only the interaction with a concave shaped CH led to larger reflected density values with respect to the incoming ones (see Figures \ref{time_dist_circle_ellip} and \ref{time_dist_convex_concave}). Due to effects of constructive interference between different reflected wave parts, in combination with a small CH density and the realistic initial wave density profile, the reflected density values can reach values more than two times larger than the initial density amplitude of the incoming wave (see, e.g., dark red structure in Figure \ref{line_profile_time_dist}, Figure \ref{time_dist_rectangular_regions}, and Table \ref{table}).
    
    \vspace{0.3cm}
    \item The interaction with a concave shaped CH is the only interaction that leads to a larger reflected density amplitude, however, the shape of the CH is not sufficient to reach such high density values. Only in combination with a realistic initial density profile (enhancement $+$ depletion) and a small CH density such strong interaction effects can be obtained. If only one of these conditions is not satisfied a large reflected density amplitude cannot be reached (see Figure \ref{CH_shapes_comparison}).

    \vspace{0.3cm}
    \item Another feature that is caused by interference effects of the reflected and the incoming wave is a strongly depleted region in the middle of the CW-CH interaction process. While a first part of the incoming wave already gets reflected and starts undergoing a phase change, the rear part of the incoming (depleted) wave still moves towards the CHB. The interference of the two depleted wave parts (reflected and incoming) leads to a depleted area that exhibits density values that are smaller than the initial depletion amplitude of the incoming wave (see, e.g., dark blue structure in Figure \ref{line_profile_time_dist} and Figure \ref{time_dist_rectangular_regions}). This also implies that for some time period during the interaction process no wave part above background density is visible. This is an important information for the analysis of reflected waves in the observations.
    \vspace{0.3cm} 
    
    \item The time-distance plots in this study show that there is not only one single reflected wave but several different reflected wave parts. What can usually be seen in the observations is the enhanced reflection which exhibits a clear and strong enhanced amplitude. However, the simulations show that an incoming wave including a realistic density profile causes several different (depleted and enhanced) reflections which together cause complex superposition effects within the density structure of the interaction process (see distribution into different regions of interest in Figure \ref{time_dist_rectangular_regions}).
    \vspace{0.3cm} 

    \item The different CH geometries lead to complex interference effects within the time-distance plots of the interaction process. These effects can on the one hand be explained by the different paths that are chosen to create these plots. On the other hand, depending on the CH density, it is possible to derive three representative incident angles, the Brewster angle, the Critical angle, and the Phase inversion angle (see Equations (\ref{thetb}), (\ref{thetc}), and (\ref{thetS})), which provide information about the transmission features and the phase characteristics of the reflected waves (see Table \ref{table2}). These properties also partially explain the interplay between the constructive and destructive interference of the involved wave parts (for more details see \citet{Piantschitsch2021}).
    \vspace{0.3cm} 

    \item By identifying and analysing different density areas within the simulation based time-distance plots (see Figure \ref{time_dist_rectangular_regions}), we can directly compare these areas and their density features to the density structure of observational time-distance plots. This enables us to derive interaction parameters from the observed interaction events which usually cannot be obtained directly from the measurements.  
    \vspace{0.3cm} 

    \end{enumerate}

    We are aware that the shapes of realistic CHs are much more complex, probably leading to different properties shown on time distance maps. This will be addressed in detail in a future study. As a next step, in the follow-up study to this paper, the simulation results obtained in this current paper, in combination with the results from \citet{Piantschitsch2023a}, will be used to partially reconstruct an actual observed CW-CH interaction event. We want to emphasize that in this theoretical study we are interested in linear and weakly non-linear CWs with a compression factor of around 1.1 or less, motivated by intensity measurements of CWs in observations \citep{Muhr2011}. The interaction results naturally change for interactions including strongly non-linear waves that lead to the evolution of shocks and other features. 

    We have to keep in mind that the simulation setup we used is still idealised, including zero gas pressure and a homogeneous background magnetic field in the $z-$direction. In future studies we plan to implement a more realistic magnetic field structure, including curved magnetic field lines at the CHB and a stratified atmosphere to be able to study vertical effects at different heights and how they influence the properties of the secondary waves. The 2D models developed in \cite{terradasetal2022,terradas2023} might be useful for these purposes. We also want to point out that in our study we used a specific inclination angle for the convex and the concave shapes of the CHs. Different choices of these inclination angles might influence the intensity of the interaction effects.

    Besides the properties of the reflected wave, the transmission coefficient and how it is related to the reflection coefficient is another important topic. Studying the properties of the transmitted wave in detail for the different CH shapes is out of the scope of this study but will be addressed in future studies. However, we would like to mention that in \citet{Piantschitsch2020} and \citet{Piantschitsch2021} we have already published theoretical results regarding the properties of the transmitted wave during a CW-CH interaction event and about how the transmission coefficient is related to the reflection coefficient.

\begin{acknowledgements}

This research was funded by the Austrian Science Fund (FWF): Erwin-Schr\"odinger fellowship J4624-N. For the purpose of Open Access, the author has applied a CC BY public copyright licence to any Author Accepted Manuscript (AAM) version arising from this submission. This work was supported by the Austrian Science Fund (FWF): I3955-N27. JT and RS acknowledge support from  the R+D+i project PID2020-112791GB-I00, financed by MCIN/AEI/10.13039/501100011033. SGH acknowledges funding by the Austrian Science Fund (FWF): Erwin-Schr\"odinger fellowship J-4560. The SDO/AIA data are available courtesy of NASA/SDO and the AIA science teams.

\end{acknowledgements}

\bibliographystyle{aa}      
\bibliography{paper_reconst_event_cit}   

\end{document}